\documentstyle[amssymb,floats,aps,prb,epsf]{revtex}
\epsfclipon 
\begin{document}
\twocolumn[\hsize\textwidth\columnwidth\hsize\csname
@twocolumnfalse\endcsname 

\draft

\title{Universal Spin Response in Copper Oxide Materials}
\author{Shiping Feng}
\address{Department of Physics, Beijing Normal University,
Beijing 100875, China \\
National Laboratory of Superconductivity, Academia Sinica,
Beijing 100080, China}
\author{Zhongbing Huang}
\address{Department of Physics, Beijing Normal University,
Beijing 100875, China}

\maketitle

\begin{abstract}
The spin response in the copper oxide materials at finite
temperatures in the underdoped and optimal doped regimes is studied
within the framework of the fermion-spin theory. The integrated
dynamical spin structure factor is almost temperature independent,
the integrated susceptibility shows the particularly universal
behavior, and the spin-lattice relaxation time is weakly
temperature dependent, which are consistent with experiments and
numerical simulations.
\end{abstract}
]

\narrowtext

After ten years of intense experimental and theoretical studies
of the copper oxide superconductors, there is now a consensus that
these materials should be described as strongly correlated electron
systems \cite{n1}, since all of the copper oxide materials have in
common the existence of a perovskite parent compound which is
insulating and has the antiferromagnetic long-range-order (AFLRO),
and changing the carrier concentration by ionic substitution or
increase of the oxygen content turns these compounds into
correlated metals leaving short range antiferromagnetic
correlations still intact \cite{n1,n2}. The short range
antiferromagnetic correlations results  in several peculiar
physical properties of the copper oxide materials: they are
responsible for the nuclear magnetic resonance (NMR) and nuclear
quadrupole resonance (NQR), and especially for the temperature
dependence of the spin-lattice relaxation rate \cite{n1,n3}. A
series of the neutron-scattering measurements \cite{n4,n5,n6} on
the copper oxide materials La$_{2-x}$Sr$_{x}$CuO$_{4}$ and
YBa$_{2}$Cu$_{3}$O$_{6+x}$ show that there is an anomalous
temperature $T$ dependence of the spin fluctuations near the
antiferromagnetic zone center in the underdoped and optimal doped
regimes, and the low-frequency dynamical susceptibility in the
optimal doped regime follows a surprisingly simple scaling
function as
$\chi^{\prime\prime}(\omega)\propto {\rm arctan}(\omega/T)$.
The NMR and NQR spin-lattice relaxation time $T_{1}$ is weakly
$T$ dependent. These unusual magnetic properties of the copper
oxide materials suggest that the normal-state can not be
described by the conventional Fermi-liquid theory.

As emphasized by many researchers \cite{n7}, the essential physics
of the copper oxide materials is contained in the doped
antiferromagnet, which may be effectively described by the
two-dimensional (2D) $t$-$J$ model acting on the space with no
doubly occupied sites. The $t$-$J$ model is reduced as the
Heisenberg model in the undoped case. In spite of its simple form
the $t$-$J$ model proved to be very difficult to analyze,
analytically as well as numerically, because of the electron
single occupancy on-site local constraint. The local nature of the
constraint is of prime important, and its violation may lead to
some unphysical results \cite{n8}. Recently a fermion-spin theory
based on the charge-spin separation is proposed \cite{n9,n10} to
incorporated this constraint, where the electron on-site local
constraint for single occupancy is satisfied even in the mean-field
approximation (MFA). The magnetic in the undoped parent compounds
is now quite well understood \cite{n1}: here, the system of
interacting localized Cu$^{2+}$ spins is well described by the 2D
Heisenberg model, and then it is clearly of great interest to
investigate in detail the crossover from the rather conventional
local moment system at zero doping to the electronic state that
forms the basis for the high-temperature superconductivity.
Therefore in this paper, we only study the spin dynamics of the
copper oxide materials within the fermion-spin theory in the
underdoped and optimal doped regimes. According to the fermion-spin
formulism \cite{n9,n10}, the electron operators can be decomposed
as $C_{i\uparrow}=h^{\dagger}_{i}S^{-}_{i}$ and
$C_{i\downarrow}=h^{\dagger}_{i}S^{+}_{i}$, with the spinless
fermion operator $h_{i}$ keeps track of the charge (holon), while
the pseudospin operator $S_{i}$ keeps track of the spin (spinon).
Within the fermion-spin theory, it has been shown \cite{n11} that
AFLRO vanishes around doping $\delta =5\%$ for the reasonable
value of the parameter $t/J=5$. The mean-field theory in the
underdoped and optimal doped regimes without AFLRO has been
developed \cite{n10}, where the mean-field order parameters
are defined  as $\chi=\langle S_{i}^{+}S_{i+\eta }^{-}\rangle =
\langle S_{i}^{-}S_{i+\eta}^{+}\rangle$, $\chi_{z}=\langle
S_{i}^{z}S_{i+\eta }^{z}\rangle$, $C=(1/Z^{2})
\sum_{\eta ,\eta ^{\prime }}\langle S_{i+\eta }^{+}
S_{i+\eta ^{\prime}}^{-} \rangle$, $C_{z}=(1/Z^{2})
\sum_{\eta ,\eta ^{\prime }}\langle S_{i+\eta }^{z}
S_{i+\eta ^{\prime }}^{z}\rangle$, and $\phi =\langle
h_{i}^{\dagger}h_{i+\eta }\rangle$ with
$\hat{\eta }=\pm \hat{x},\pm \hat{y}$, and $Z$ is the number of
nearest neighbor sites. In this case, the low-energy behavior can
be described \cite{n10} by the effective Hamiltonian
$H=H_t+H_J$ with
\begin{mathletters}
\begin{eqnarray}
H_t&=&-t \sum_{i\eta}h_{i}h_{i+\eta}^{\dagger}(S_{i}^{+}
S_{i+\eta}^{-}+S_{i}^{-}S_{i+\eta}^{+})+h.c. \nonumber \\
&+& \mu \sum_{i}h_{i}^{\dagger}h_{i} , \\
H_J&=&J_{eff}\sum_{i\eta}[{1 \over 2}(S_{i}^{+}
S_{i+\eta}^{-}+S_{i}^{-}S_{i+\eta}^{+})+S_{i}^{z}S_{i+\eta}^{z}],
\end{eqnarray}
\end{mathletters}
where $J_{eff}=J[(1-\delta )^2-\phi ^2]$, and $\mu$ is the chemical
potential which enforce $\langle h_{i}^{\dagger}
h_{i}\rangle=\delta$.

In the framework of the charge-spin separation, the basic
low-energy excitations are holons and spinons. The charge dynamics
can be discussed based on the Ioffe-Larkin combination rule
\cite{n12}, however, since the spin fluctuations couple only to
spinons, and therefore no composition law is required in discussing
the spin dynamics \cite{n12}, but the strongly correlation between
holons and spinons still is considered through the holon's order
parameters $\phi$ entering in the spinon propagator, which means
that the spinon moves in the background of holons, and the cloud
of distorted holon background is to follow spinons, therefore the
dressing of the spinon by holon excitations is the key ingredient
in the explanation of the spin dynamics. According to Ioffe-Larkin
combination rule \cite{n12}, we \cite{n13} have discussed the
optical conductivity, Drude weight, and resistivity of the copper
oxide materials in the underdoped and optimal doped regimes by
considering fluctuations around the mean-field solution, where the
dominant dynamical effect is due to the strongly spinon-holon
interaction in Hamiltonian (1). We believe that this strongly
spinon-holon interaction also will dominate the spin dynamics
within the same regimes. The mean-field spinon Green's functions
$D^{(0)}(k,\omega)$ and $D^{(0)}_{z}(k,\omega)$ and mean-field
holon Green's function $g^{(0)}(k,\omega)$ have been given in
Ref. \cite{n10}. In this paper, we limit the holon part to the
first-order (mean-field level) since some physical properties can
be well described at this level \cite{n10}, and spin fluctuations
couple only to spinons as mentioned above. However, the
second-order correction for the spinon is necessary for the
discussion of the spin dynamics. The second-order spinon
self-energy diagram from the holon pair bubble is shown in Fig. 1.
Since the spinon operators obey the Pauli algebra, we map the
spinon operator into the spinless-fermion representation in terms
of the 2D Jordan-Wigner transformation \cite{n14} for the formal
many particle perturbation expansion. After then the spinon
Green's function in the spinon self-energy diagram shown in
Fig. 1 is replaced by the mean-field spin Green's function
$D^{(0)}(k,\omega)$. In this case, we obtain the second-order
spinon self-energy as,
\begin{eqnarray}
\Sigma_{s}^{(2)}(k,\omega)&=&-(Zt)^{2}{1\over N^2}\sum_{pp'}
(\gamma_{k-p}+\gamma_{p'+p+k})^{2}B_{k+p'} ~~~~~\nonumber \\
&\times& \left ({F_{1}(k,p,p')\over \omega +\xi_{p+p'}-\xi_{p}+
\omega_{k+p'}+i0^{+}} \right. \nonumber \\
&-&\left. {F_{2}(k,p,p')\over \omega +\xi_{p+p'}-\xi_{p}-
\omega_{k+p'}+i0^{+}} \right ) ,
\end{eqnarray}
where $F_{1}(k,p,p')=n_{F}(\xi_{p+p'})[1-n_{F}(\xi_{p})]+[1+n_{B}
(\omega_{k+p'})][n_{F}(\xi_{p})-n_{F}(\xi_{p+p'})]$, $F_{2}(k,p,p')
=n_{F}(\xi_{p+p'})[1-n_{F}(\xi_{p})]-n_{B}(\omega_{k+p'})[n_{F}
(\xi_{p})-n_{F}(\xi_{p+p'})]$,
$\gamma_{k}=(1/Z)\sum_{\eta}e^{ik\cdot \hat{\eta}}$,
$\epsilon =1+2t\phi/J_{eff}$, $B_k=ZJ_{eff}[(2\epsilon \chi_z+\chi)
\gamma_{k}-(\epsilon \chi +2\chi_z)]/\omega_{k}$, $n_{F}(\xi_{k})$
and $n_{B}(\omega_{k})$ are the Fermi and Bose distribution
functions, respectively, the mean-field holon excitation spectrum
$\xi_{k}=2Z\chi t\gamma_{k}+\mu$, and the mean-field spinon
excitation spectrum $\omega_{k}$ is given in Ref. \cite{n10}. Then
the full spinon Green's function is obtained as
$D^{-1}(k,\omega)=D^{(0)-1}(k,\omega)-\Sigma_{s}^{(2)}(k,\omega)$.
Since the local constraint of the $t$-$J$ model has been treated
exactly in the previous mean-field theory \cite{n10}, and it is
natural satisfied in the above perturbation expansion based on this
mean-field theory.

We are now ready to discuss the spin dynamics. The dynamical spin
response, as manifested by the dynamical spin structure factor
$S(k,\omega)$ and the susceptibility $\chi(k,\omega)$, are given as
$S(k,\omega)=Re\int^{\infty}_{0}dt e^{i\omega t}\langle S^{+}_{k}
(t)S^{-}_{k}(0)\rangle=2{\rm Im}D(k,\omega)/(1-e^{-\beta \omega})$
and $\chi^{\prime\prime}(k,\omega)=(1-e^{-\beta \omega})S(k,\omega)
=2{\rm Im}D(k,\omega)$. The properties of $S(k,\omega)$ and
$\chi^{\prime\prime}(k,\omega)$ in different $k$ directions have
been discussed \cite{n15}, and the results showed that there is the
anomalous temperature $T$ dependence of the spin fluctuations near
the antiferromagnetic point $Q=(\pi,\pi)$. In this paper we are
interested in the universal behavior of the integrated dynamical
response. The integrated dynamical spin structure factor and
integrated susceptibility are expressed as,
\begin{eqnarray}
\bar{S} (\omega)&=&S_{L}(\omega)+S_{L}(-\omega)=
(1+e^{-\beta \omega})S_{L}(\omega), \nonumber \\
S_{L}(\omega)&=&{1\over N}\sum_{k}S(k,\omega),
\end{eqnarray}
and
\begin{eqnarray}
I(\omega, T)={1\over N}\sum_{k}\chi^{\prime\prime}(k,\omega),
\end{eqnarray}
respectively. We have performed a numerical calculation for the
integrated spin structure factor (3) and integrated susceptibility
(4). The result of the integrated spin structure factor for the
parameter $t/J=2.5$ with the temperature $T=0.3J$ (solid line),
$T=0.4J$ (dashed line), and $T=0.5J$ (dotted line) at the doping
(a) $\delta =0.08$, and (b) $\delta=0.15$ is plotted in Fig. 2.
From Fig. 2, it is shown that the integrated spin structure factor
is almost temperature independent and the shape appears to be
particularly universal in the underdoped and optimal doped
regimes. \=S$(\omega)$ is decreased with increasing energies for
$\omega <0.5t$, and almost constant for $\omega \geq 0.5t$, which
is consistent with the experiments \cite{n16} and numerical
simulations \cite{n17}. In correspondence with the integrated spin
structure factor, the result of integrated susceptibility at the
doping $\delta = 0.15$ for the parameter $t/J=2.5$ with the
temperature $T=0.2J$ (solid line), $T=0.3J$ (dashed line), and
$T=0.4J$ (dotted line) is plotted in Fig. 3. For comparison, the
function $b_{1}{\rm arctan}[a_{1}\omega/T+a_{3}(\omega/T)^{3}]$
with $b_{1}=0.23$, $a_{1}=2.0$, and $a_{3}=1.4$ is also plotted
in Fig. 3 (dot-dashed line). Our results show that the integrated
susceptibility is almost constant above $\omega/T >1$ and then
begin to decrease with decreasing $\omega/T$ for $\omega/T <1$.
It is quite remarkable that our theoretical results of the
integrated susceptibility are scaled approximately as
$I(\omega,T)\propto {\rm arctan}[a_{1}\omega/T+a_{3}
(\omega/T)^{3}]$, which is in very good agreement with the
experiments \cite{n16}.

\begin{figure}[prb]
\epsfxsize=3.0in\centerline{\epsffile{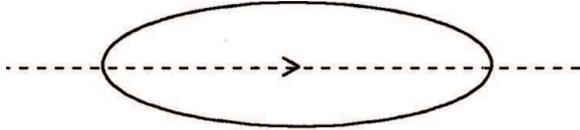}}
\caption{The spinon's second-order self-energy diagram. The
solid and dashed lines correspond to the holon and spinon
propagators, respectively.}
\end{figure}

\begin{figure}[prb]
\epsfxsize=3.0in\centerline{\epsffile{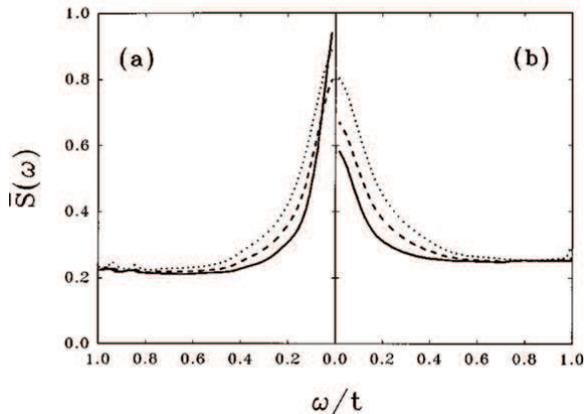}}
\caption{The integrated dynamical spin structure factor at the
doping (a) $\delta=0.08$ and (b) $\delta=0.15$ for $t/J=2.5$ with
the temperature $T=0.3J$ (solid line), $T=0.4J$ (dashed line), and
$T=0.5J$ (dotted line).}
\end{figure}

\begin{figure}[prb]
\epsfxsize=3.0in\centerline{\epsffile{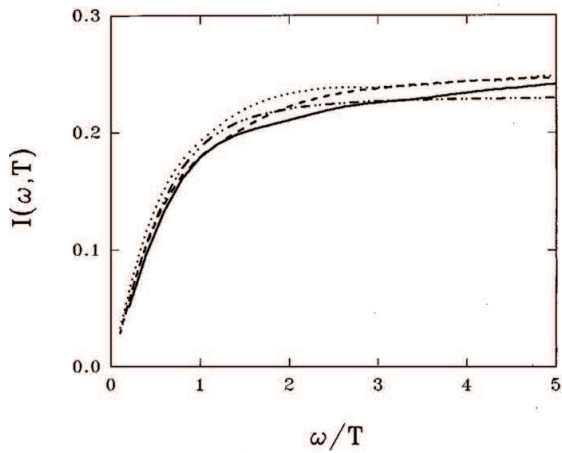}}
\caption{The integrated susceptibility at the doping $\delta=0.15$
for $t/J=2.5$ with the temperature $T=0.2J$ (solid line), $T=0.3J$
(dashed line), and $T=0.4J$ (dotted line). The dot-dashed line is
the function $b_{1}{\rm arctan}[a_{1}\omega/T+a_{3}(\omega/T)^{3}]$
with $b_{1}=0.23$, $a_{1}=2.0$, and $a_{3}=1.4$.}
\end{figure}

\begin{figure}[prb]
\epsfxsize=3.0in\centerline{\epsffile{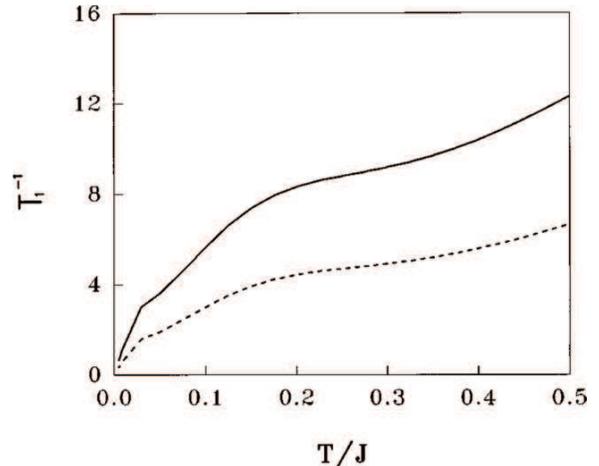}}
\caption{The spin-lattice relaxation time at the doping
$\delta=0.15$ with $t/J=2.5$ for the field applied parallel to
C axis (solid line) and perpendicular to C axis (dashed line).}
\end{figure}

The temperature dependence of the susceptibility converges to a
universal function of $\omega/T$ is very significant because of
its relation to other normal state properties, such as the
temperature dependences of the spin-lattice relaxation time. The
NQR spin-lattice relaxation time $T_{1}$ is expressed as,
\begin{eqnarray}
{1\over T_{1}}={2K_{B}T\over g^{2}\mu^{2}_{B}\hbar}
\lim\limits_{\omega\rightarrow 0}{1\over N}\sum_{k}
F^{2}_{\alpha}(k){\chi^{\prime\prime}(k,\omega)\over \omega},
\end{eqnarray}
where $g$ is the $g$ factor, $\mu_{0}$ is the Bohr magneton, and
the form factors $F_{\alpha}(k)=(F_{\bot}(k)$, $F_{\Vert}(k))$,
with $F_{\bot}(k)$ and $F_{\Vert}(k)$ are for the field applied
parallel and perpendicular to the C axis, respectively. The form
factors have dimension of energy, and magnitude determined by
atomic physics, and $k$ dependence determined by geometry. For the
comparison with experiments, the form factors $F_{\bot}(k)$ and
$F_{\Vert}(k)$ are chosen as proposed in Ref. \cite{n18}. The
spin-lattice relaxation time $T_{1}$ in Eq. (5) has been evaluated
numerically and the results for $t/J=2.5$ with the doping
$\delta =0.15$ for the field applied parallel to C axis (solid
line) and perpendicular to C axis (dashed line)  are plotted in
Fig. 4, where we have chosen units $\hbar=K_{B}=1$. From Fig. 4,
it is shown that $T_{1}$ is very weakly dependent on T in the
optimal doped regime. Some experiments \cite{n5} show that
$1/T_{1}$ approaches nearly the temperature independent at high
temperature for La$_{2-x}$Sr$_{x}$CuO$_{4}$, and the weakly
temperature dependent for YBa$_{2}$Cu$_{3}$O$_{7}$, in the optimal
doped regime. Although the simplest $t$-$J$ model can not be
regarded as the complete model for the quantitative comparison
with the copper oxide materials, but our results are in qualitative
agreement with these remarkable experiments \cite{n5}.

In the fermion-spin theory, the charge and spin degrees of freedom
of the physical electron are separated as the holon and spinon,
respectively. Although both holons and spinons contributed to the
charge and spin dynamics, but it has been shown that the scattering
of holons dominates the charge dynamics \cite{n13}, while the
present results shows that scattering of spinons dominates the spin
dynamics. The spin dynamics probe local magnetic fluctuations and
are a very detailed and stringent test of microscopic theories. Our
theoretical results within the fermion-spin formulism leads to the
behaviors similar to that seen in the experiments and numerical
simulations. To our present understanding, the main reasons why the
present theory is successful in studying the normal-state property
of the strongly correlated copper oxide materials are that (1) the
electron single occupancy on-site local constraint is exactly
satisfied during the above analytic calculation. Since the
anomalous normal-state property of the copper oxide materials are
caused by the strong electron correlation in these systems
\cite{n1,n2,n3}, and can be effectively described by the $t$-$J$
model \cite{n7}, but the strong electron correlation in the $t$-$J$
model manifests itself by the electron single occupancy on-site
local constraint, which means that the electron Hilbert space is
severely restricted due to the strong electron repulsion
interaction. This is why the crucial requirement is to treat this
constraint exactly during the analytic discussions. (2) Since the
local constraint is satisfied even in the MFA within fermion-spin
theory, the extra gauge degree of freedom occurring in the
slave-particle approach does not appear here \cite{n9,n10}, then
spinons and holons within  the fermion-spin theory are by
themselves gauge invariant, they are real and can be interpreted
as physical excitations. In this case, there are two relaxation
times for spinons and holons, respectively, the spinon relaxation
time is responsible to the spin dynamics, while holon relaxation
time is responsible to the charge dynamics. This important issue
to the copper oxide materials within the charge-spin separation
has been emphasized by Laughlin \cite{n19}.

In summary, we have studied the spin dynamics of the copper oxide
materials in the underdoped and optimal doped regimes within the
framework of the fermion-spin theory. Our results show that the
strongly correlated renormalization effects for the spinon due to
the strongly spinon-holon interaction are very important for the
spin dynamics. The integrated dynamical spin structure factor,
integrated susceptibility, and spin-lattice relaxation time
are discussed, and the results are qualitative consistent with
the experiments and numerical simulations.

\acknowledgments
The authors would like to thank Prof. H. Q. Lin for helpful
discussions. This work is supported by the National Science
Foundation Grant No. 19474007 and the Trans-Century Training
Programme Foundation for the Talents by the State Education
Commission of China. The partial support from the China-Link
Program at the Chinese University of Hong Kong and the Earmarked
Grant for Research from the Research Grants Council of Hong Kong,
China under project CUHK 311/96P account number 2160068 are also
acknowledged.

\end{document}